\documentclass[showpacs,twocolumn,amsmath,amssymb,prb]{revtex4-1}
\usepackage{graphicx}
\usepackage{bm}
\usepackage{epsfig}
\usepackage{color}
\usepackage[]{natbib} 
\usepackage{url}

\begin{document}
\title{First-Principles Calculation of Electronic Energy Level Alignment\\ at Electrochemical Interfaces}
\author{Yavar T. Azar }
\author{Mahmoud Payami}
\email[Corresponding author: ]{mpayami@aeoi.org.ir}
\affiliation{Theoretical and Computational Physics Group, NSTRI, AEOI, P.~O.~Box~14395-836, Tehran, Iran}
\date{\today}
\begin{abstract}
Energy level alignment at solid-solvent interfaces is an important step in determining the properties of electrochemical systems. The positions of conduction and valence band edges of a semiconductor are affected by its environment. In this study, using first-principles DFT calculation, we have determined the level shifts of the semiconductors TiO$_2$ and ZnO at the interfaces with MeCN and DMF solvent molecules. The level shifts of semiconductor is obtained using the potential difference between the clean and exposed surfaces of asymmetric slabs. In this work, neglecting the effects of present ions in the electrolyte solution, we have shown that the solvent molecules give rise to an up-shift for the levels, and the amount of this shift varies with coverage. It is also shown that the shapes of density of states do not change sensibly near the gap. Molecular dynamics simulations of the interface have shown that at room temperatures the semiconductor surface is not fully covered by the solvent molecules, and one must use intermediate values in an static calculations. 
\end{abstract}
\maketitle
\section{Introduction}\label{sec1}
Abundance and ease of preparation have made titanium oxide(TiO$_2$) and zinc-oxide (ZnO) as the mostly 
used semiconductors in electrochemical applications\cite{gratzel2001photoelectrochemical,mane2005nanocrystalline}. Design and engineering of an electrochemical system 
need a deep understanding of the semiconductor behaviors at its interface with other materials.

The alignment of electronic energy levels at solid-solvent interfaces plays a crucial role in the functionality of electrochemical systems\cite{ishii1999energy,greiner2012universal,narioka1995electronic,cheng2012alignment,Kharche2014}. 
Hydrogen-evolution in photo-electrochemical cells\cite{morales2014nanostructured,laursenelectrochemical} and electron injection in dye-sensitized solar cells (DSSC)\cite{o1991low} are two well-known instances of reactions which highly depend on the relative alignment of electronic energy levels at the semiconductor-solvent interface. 

In a realistic electrochemical system, the relative positions of conduction and valence band edges (CB and VB) of the semiconductor are sensitive to its environmental factors: pH of the solvent, the concentration of dissolved ions, defects and adsorbents at the semiconductor surface. 
On the other hand, the short- and long-range interactions of the solvent molecules with the semiconductor surface atoms could significantly affect the electronic structure of the semiconductor\cite{Kharche2014}.

Adsorption of solvent molecules on the semiconductor surface results in a net electric dipole (see appendix~\ref{appsec1}), which in 
turn, causes shifts in the energy positions of the conduction and valance bands at the semiconductor 
surface\cite{mosconi2012solvent}. To explore the interrelation between the molecular structure of the solvent and the level 
alignment at the interface, one has to consider the reasons behind the formation of 
electric dipole 
layer at the surface\cite{cheng2012alignment}. 
This surface dipole moment results from the permanent dipoles of adsorbate 
molecules and the charge displacements through the formation of chemical bonding between semiconductor 
surface and the solvent molecules\cite{kera2004impact}. 

The description of detailed mechanisms involved at interfaces is not possible unless one uses computational methods. The density functional theory (DFT)\cite{HK64,KS65} and its time-dependent extension (TDDFT)\cite{RG84} have played significant roles in the design and engineering of electrochemical systems among which hydrogen evolution systems\cite{hinnemann2005biomimetic,greeley2006computational} and DSSC's\cite{sumita2011water,de2007time,mathew2014dye,azar2014efficiency,azar2015theoretical} are the most common examples.     

In this work, using first-principles DFT calculations, we aim to explore how the changes in semiconductor 
environment, such as molecular structure of the solvent and its surface coverage, could affect the 
energy levels at the interface. Here we consider the two most popular solvents, acetonitrile (MeCN) and dimethylformamide (DMF). 

It should be mentioned that in an electrochemical system, one needs to identify the exact positions\cite{Kharche2014} of CB and VB, which demands some sophisticated methods such as GW\cite{hedin1965new,aulbur1999quasiparticle} or others\cite{georges1996dynamical,tran2009accurate}. 
However, since these positions for TiO$_2$ and ZnO are well-known, and in this work we aim to determine just the amounts of shifts for the levels, regardless of the absolute positions, we do not need to perform such relative expensive calculations, and therefore the results obtained based on KS-DFT method suffices. 

In the calculations, we have considered adsorption geometry and binding energy as a function of surface coverage, and obtained a non-linear dependence for the binding energy and level shifts.  

Separating the total dipole moment of the interface into different components (appendix~\ref{appsec1}) showed that the dominant charge transfer to the interface region is from the adsorbed solvent molecules which causes formation of a net dipole towards the surface, and this, in turn, results in an upward level shifts for the semiconductor.    

In our periodic slab scheme of calculations, using the dipole correction method\cite{Bengtsson1999,Meyer2001}, we have determined the potential difference between the clean and exposed surfaces, and used it to estimate the level shifts of the semiconductor in presence of solvent molecules.

Finally, using {\it ab-initio} molecular dynamics (MD) for the time-evolution of solvent-semiconductor interface, we have shown that 
the thermal effects and interaction of solvent molecules prevent the formation of a full coverage for semiconductor.

The structure of this paper is as follows. In section 2 we present the computational details; the calculation results are presented and discussed in section 3; and we conclude this work in section 4. Finally, the relations for displacement charge and surface dipole moment are derived in appendix~\ref{appsec1}. 

\section{Computational Methods}
Modeling the titania surface, we have constructed an anatase 8-(TiO$_2$)-layer slab with (101) surface using a $3\times 1$ supercell along [010] and [10$\bar{1}$] directions. For zinc oxide surface, we have made an 8-(ZnO)-layer slab with (10$\bar{1}$0) surface using $3\times 2$ supercell along [000$\bar{1}$] and [11$\bar{2}$0] directions. In all of our calculations the systems are fully relaxed. It has been shown by others\cite{martsinovich2010electronic} that taking an 8-layer slab is suitable for low-index anatase surface calculations, and the effects of relaxations on electronic structure has also been studied by other researchers\cite{lazzeri2001structure,zhao2010surface}.  

In this work, all the electronic-structure calculations are based on the DFT and the self-consistent solution of the Kohn-Sham (KS) equations\cite{KS65} using the Quantum ESPRESSO (QE) code package\cite{QE-2009} within the PBE generalized gradient approximation\cite{PBE96} for the XC energy functional. For the atoms Zn, Ti, O, C, N, and H  we have used the ultrasoft pseudopotentials Ti.pbe-sp-van\_ak.UPF, O.pbe-van\_bm.UPF, C.pbe-van\_bm.UPF, N.pbe-van\_ak.UPF, S.pbe-van\_bm.UPF, F.pbe-n-van.UPF, and H.pbe-van\_ak.UPF available at http://www.quantum-espresso.org. The kinetic-energy cutoff for the plane-wave basis set were chosen 28 and 220 Ry for the wave functions and charge density, respectively. For the Brillouin-zone integrations, a $2\times2\times1$ grid was used. 

Choosing a reasonable reference potential is a crucial step in the study of level alignment because, the comparison of surface levels before and after the adsorption needs a unique reference point. One usual method in models with slab geometry  is plotting the planar average potential as a function of $z$ which is normal to the surface, and choosing the reference potential, $V(\infty)$, at a point far from surfaces in vacuum region. The plane-averaged potential is defined as:
  
  \begin{equation}\label{eq1}
  \bar{V}(z)=\frac{1}{S}\int_{\rm supercell}^{}V(x,y,z)dxdy,
  \end{equation}
where $S$ is the surface area of the  supercell.

Identification of the reference point for a symmetric slab is trivial because the average potential becomes flat deep inside the vacuum region. However, for asymmetric slabs, because of a net surface dipole moment, the electrostatic potentials are different in the vacua at two sides of the slab. Solution of the Poisson's equation with a periodic boundary condition introduces an artificial uniform electric field across the supercell which deteriorates the flatness of potentials at the two sides of the slab. In this case we have no well-defined reference point as in the symmetric slab case. One workaround is to take much thicker symmetrized slabs which is computationally expensive. A more convenient way is to use the dipole correction method\cite{Bengtsson1999,Meyer2001} in which the artificial linear potential is compensated by a sawtooth-like external potential. In this work we have used the second method as implemented in QE code package. 

The Born-Oppenheimer {\it ab initio} molecular-dynamics simulations at room temperature are performed employing the SIESTA code package\cite{soler2002siesta} within the DFT and the PBE\cite{PBE96}  
level of approximation for the exchange-correlation. For basis sets, a split-valence 
double-$\zeta$ basis augmented by polarization functions (DZP) are used along with the nonrelativistic pseudopotentials for all atoms. Real-space integrals were performed on a mesh with a 150 Ry cutoff.
% % % % % % % % % % % % % % % % % % % % % % % % % % % % % % % % % % % % % % % % % % %
\section{Results and Discussions}
\begin{figure}
	\centering
	\includegraphics[width=0.7\linewidth]{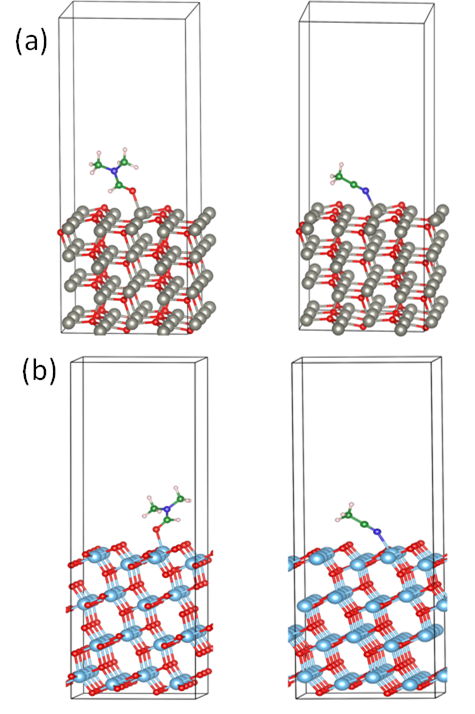}
	\caption{Equilibrium geometries of adsorbed DMF (left) and MeCN (right) on (a)- the ($10\bar 10$) surface of ZnO, and (b)- the (101) surface of TiO$_2$.}
	\label{fig1}
\end{figure}

The equilibrium geometries and electronic structure of slab-solvent interfaces are calculated using different coverages of 1, 2, 3, 6 solvent molecules per supercell for both TiO$_2$ and ZnO. The resulting equilibrium geometries of an MeCN and a DMF adsorbed on ZnO and TiO$_2$ surfaces are compared in Fig.\ref{fig1}.
As is seen from the figure, the orientations of each molecule on the TiO$_2$ and ZnO surfaces are more or less the same, perhaps because of the fact that both Ti and Zn are transition metals which are surrounded by O atoms in a similar manner. 

\begin{table}[]\small
\caption{Binding energies, $E_{\rm b}$, and electrostatic potential differences between two sides of slabs, $\Delta V$, in electron-volts, molecule-surface distance, $d_1$, and selected intra-molecular distance, $d_2$, in $\AA$, and the interface dipoles, $\mu$, in atomic units for different coverages, n.}
 \centering
\resizebox{\columnwidth}{!} 
{
\begin{tabular}{lllllll}
\hline	 System    & n     & E$_b$(eV) & $\Delta V$(eV)  & d$_1$($\AA$)  & d$_2$$(\AA)$ & $\mu$(a.u.) \\ 
\hline	 n-MeCN/ZnO & 1     & 0.693     & 1.30        & 2.08 & 1.16 & 1.48  \\ 
		            & 2     & 0.660     & 2.05        & 2.10 & 1.16 & 2.27  \\ 
		            & 3	    & 0.614     & 2.41        & 2.10 & 1.16 & 2.65  \\ 
		            & 6	    & 0.387     & 3.70        & 2.24 & 1.16 & 4.09  \\ 
		&    &       &             &      &      &    \\ 
		n-DMF/ZnO	& 1		& 0.790     & 1.60        & 2.05 & 1.26 &   1.76 \\ 
		            & 2		& 0.733     & 2.43        & 2.09 & 1.26 &   2.71 \\ 
		            & 3		& 0.709     & 2.88        & 2.13 & 1.26 &   3.20 \\ 
		            & 6		& 0.468     & 4.06        & 2.20 & 1.26 &   4.50 \\    
		&     &   &      &      &      &    \\ 
		n-MeCN/TiO$_2$ & 1  & 0.594     & 1.21 & 2.28 & 1.17 &   1.39 \\ 
		               & 2  & 0.555     & 1.95 & 2.29 & 1.17 &   2.24 \\ 
		               & 3	& 0.485     & 2.57 & 2.30 & 1.16 &   2.97 \\ 
		               & 6	& 0.341     & 3.41 & 2.41 & 1.16 &   3.95 \\ 
		&      &     &      &      &      &   \\ 
		n-DMF/TiO$_2$  & 1	& 0.641     & 1.92 & 2.18 & 1.25 &   2.21 \\ 
		               & 2	& 0.500     & 2.93 & 2.35 & 1.25 &   3.39 \\ 
		               & 3	& 0.449     & 3.84 & 2.30 & 1.24 &   4.42 \\ 
		               & 6	& 0.327     & 5.23 & 2.41 & 1.23 &   6.05 \\ 
		\hline
	\end{tabular}
}
	\label{tab1}
\end{table}

Geometrical parameters and binding energies for different coverage values are calculated and tabulated 
in Table~\ref{tab1}. 
As is seen from table, with increasing the coverage, the distance between the solvent molecule and surface, ($d_1$), increases which is consistent with behavior of the binding energy, $E_b$. The relatively high values for the binding energies imply that the  bondings between the solvent molecules and the surfaces have near covalent character. As to the ``selected intra-molecular distance'', $d_2$, it is defined as the distance between the two nearest atoms of the solvent molecule to the surface of the semiconductor. With decreasing the binding energy, we expect that $d_2$ decreases down to its value for isolated molecule (1.23~$\AA$ for DMF, and 1.16~$\AA$ for MeCN) and this change is so small that they are more or less constant. 

% % % % % % % % % % % % % % % % % % % % % % % % % % % % % % % % % % % % % % % %
% % % % % % % % % % % % % % % % % % % % % % % % % % % % % % % % % % % % % % % % % %

To gain insight into the solvent-surface interactions, we have examined the charge redistribution between molecule and surface after the bond formation. Considering the periodic geometry of interface in $(x,y)$ plane, and averaging the density over this plane, the resulting laterally averaged charge density: 
\begin{equation}\label{eq2}
\bar{\rho}(z)= \int \rho({\bf r}) dx dy
\end{equation}
would be an appropriate quantity for more detailed analysis of charge displacement.

\begin{figure}
\centering
\includegraphics[width=\linewidth]{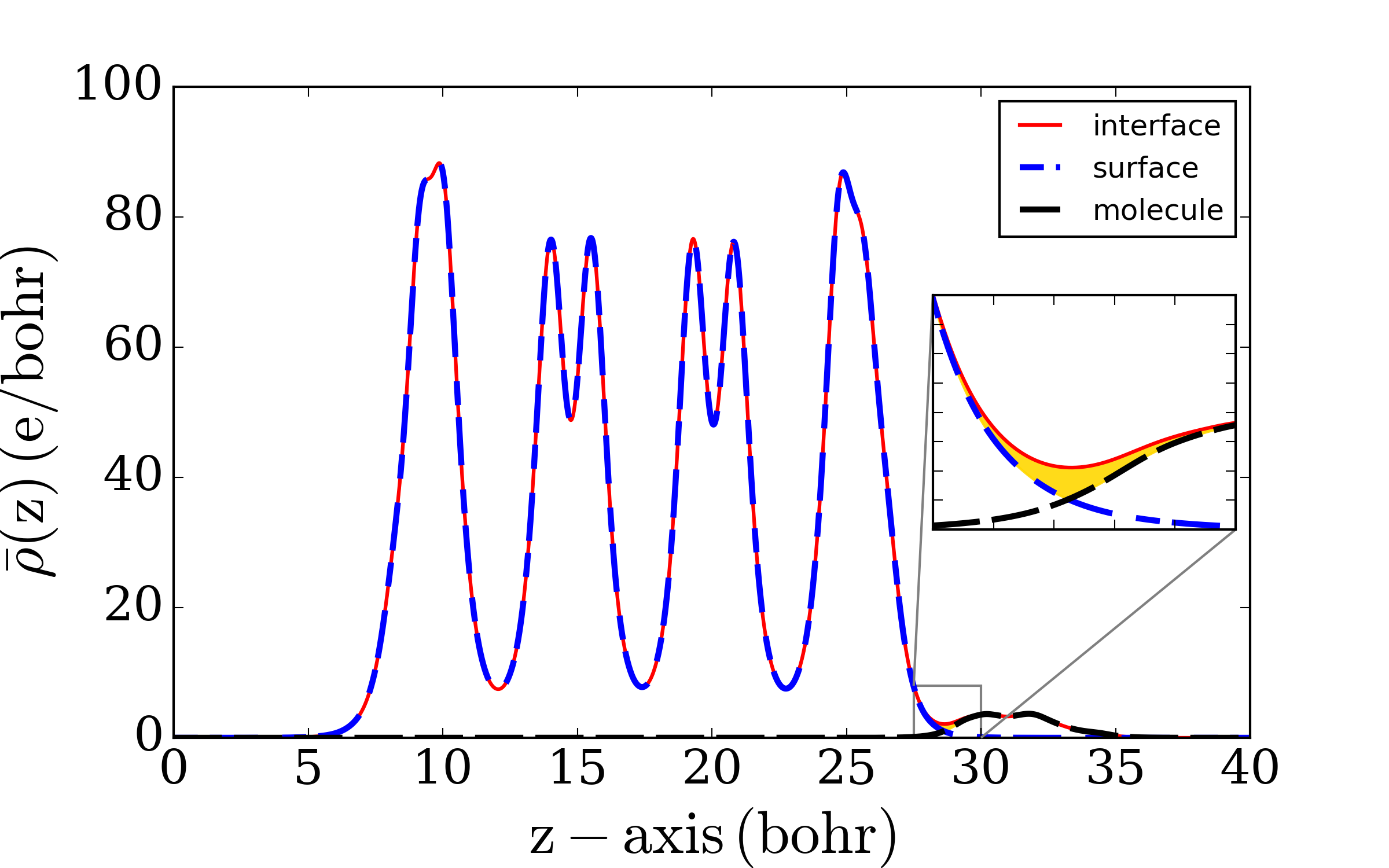}
\caption{Averaged charge density for MeCN/ZnO after adsorption (thin solid red line) as well as for the isolated molecule (dashed black line) and surface (dashed blue line) at the same relative positions. The nearly full coincidence of the three plots indicates that there is no significant charge redistribution far from the interface region. The yellow region in the zoomed part represents the amount of injected charge. }
\label{fig2}
\end{figure}

We have calculated this averaged quantity for MeCN/ZnO after adsorption as well as for the isolated molecule and surface at the same relative ionic positions, and have shown the result in Fig.~\ref{fig2}. From this figure we see that there is some charge injection from the molecule into the interface region. 

% % % % % % % % % % % % % % % % % % % % % % % % % % % % % % % % % % %
% % % % % % % % % % % % % % % % % % % % % % % % % % % % % % % % % % % % % % % %

\begin{figure}
\centering
\includegraphics[width=0.98\linewidth]{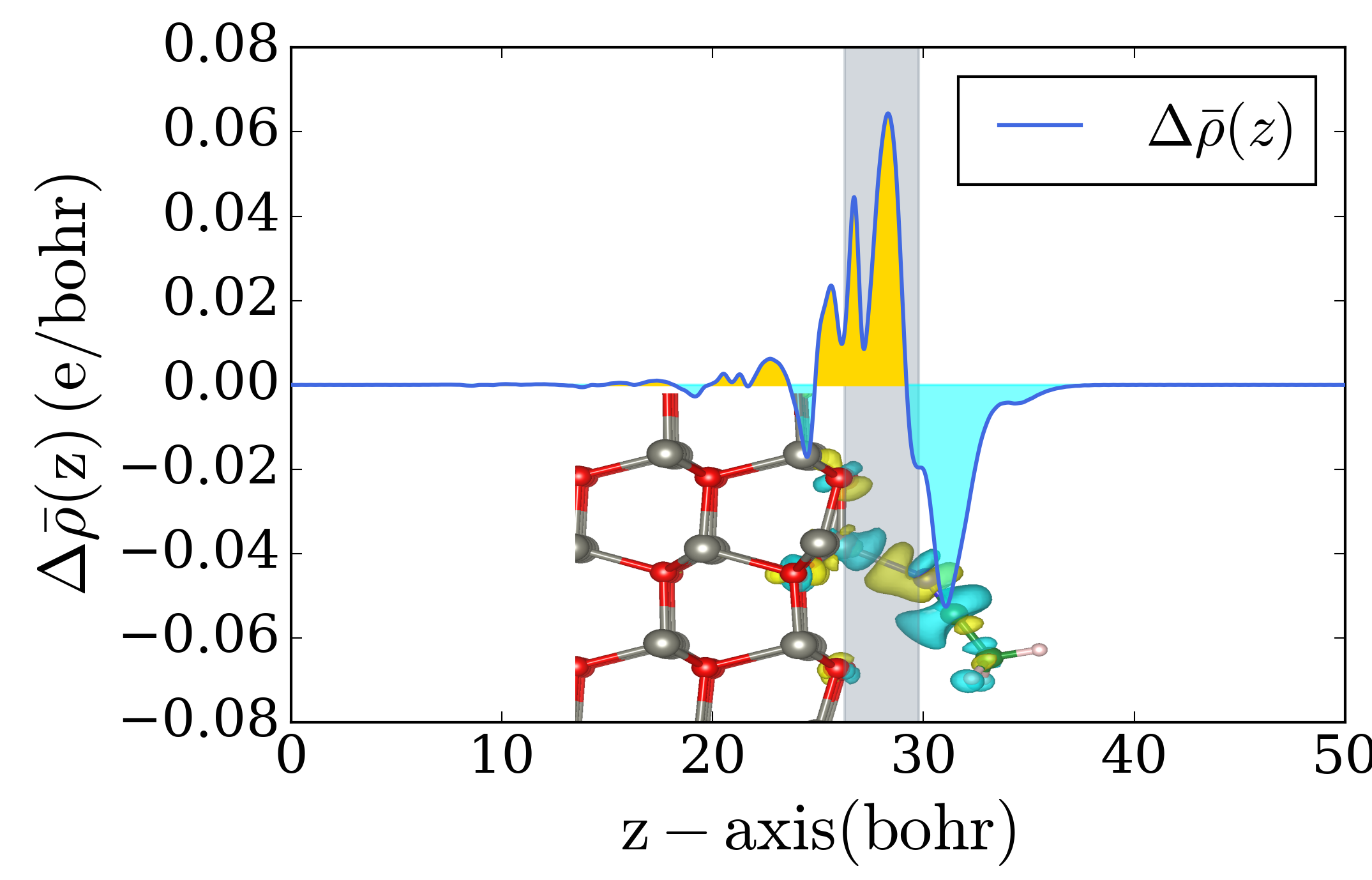}
\caption{Average charge displacement along normal direction and 3D charge difference isosurface. The grey interval specifies the interface region. Yellow and cyan regions represent increase and decrease of charge, respectively.}
\label{fig3}
\end{figure}

The charge displacement is better presented when we consider the electronic charge-difference quantity defined by (appendix~\ref{appsec1}):

% % % % % % % % % % % % % % % % % % % % % % % % % % % % % % % % %
\begin{equation}\label{eq3}
 \Delta \rho({\bf r}) = \rho_{\rm comb}({\bf r}) - \rho_{\rm sur}({\bf r}) -\rho_{\rm mol}({\bf r})
\end{equation}
% % % % % % % % % % % % % % % % % % % % % % % % % % % % % % % % % % % % % %
where the first, second, and third terms on the right refer to the electronic densities of the combined molecule-surface system, deformed surface, and deformed molecule, respectively. Commonly the ionic positions are chosen in such a way that there is no contributions from the positive ions in the charge transfer. 
The average of this quantity over the $(x,y)$ plane, is shown in Fig.~\ref{fig3} as a function of $z$ which is normal to the surface. We note that there is a charge increment in the interface region. The interface region boundaries are defined by two parallel ($x,y$)-planes, one containing the outermost atom of the surface and another containing the nearest atom of the molecule to the surface. Since on the one hand the charge increment resides in the region between these two boundary atoms, and on the other hand the binding energy is of the order of a covalent bonding between two atoms, we may interpret the bonding between the solvent molecule and the surface as having a near covalent character. 

As one notes in Table~\ref{tab1}, the binding energy per molecule decreases with increasing the coverage, leading to weaker bondings of molecules to the surface. Also it is evident from Fig.~\ref{fig3} that the adsorbed molecule has more contribution in charge displacement than the surface, which leads to the formation of a dipole layer near the surface. Since the orientation of generated dipole is towards the surface, we expect an up-shift for the semiconductor energy levels.   

Using the splitting of electronic charge density as in Eq.~(\ref{eq3}), the $z$-component of dipole moment of the combined system can be written as (appendix~\ref{appsec1}):

\begin{equation}\label{eq4}
\mu_{\rm comb}=\mu_{\rm mol}+\mu_{\rm sur}+\mu_{\rm chem}
\end{equation}
where $\mu_{\rm chem}$ is the dipole originating from the charge displacement (chemical bonding). The values of $\mu_{\rm comb}$, $\mu_{\rm mol}$, $\mu_{\rm sur}$, and $\mu_{\rm chem}$ in the MeCN/ZnO system are calculated to be 1.46, 0.81, 0.14, and 0.51 a.u., respectively. These values imply that the charge displacement has a significant contribution (after the molecule itself) in the total dipole moment. 
  
In the next step, we have studied how the adsorption of solvent molecules on the surface affects the potential difference between the two sides of the slab. For this purpose, we have used the calculated averaged electrostatic potential after applying the dipole correction\cite{Bengtsson1999}:

\begin{equation}\label{eq5}
V_{\rm dip}(z)=4\pi\frac{\mu}{S} \left(\frac{z}{z_m}-\frac{1}{2}\right) ,\;\;\;\; 0 < z < z_m
\end{equation}
where $\mu$, $S$, and $z_m$ are the dipole of the supercell, surface area of the supercell, and the length of the region over which the correction has been applied, respectively.
Equation~(\ref{eq5}), leads to a potential jump across the slab:
\begin{equation}\label{eq6}
\Delta V=4\pi\frac{\mu}{S},
\end{equation} 
which we could interpret as the difference between the asymptotic potentials of clean surface, $V(-\infty)$, and exposed surface, $V(+\infty)$.

\begin{figure}
\centering
\includegraphics[width=0.8\linewidth]{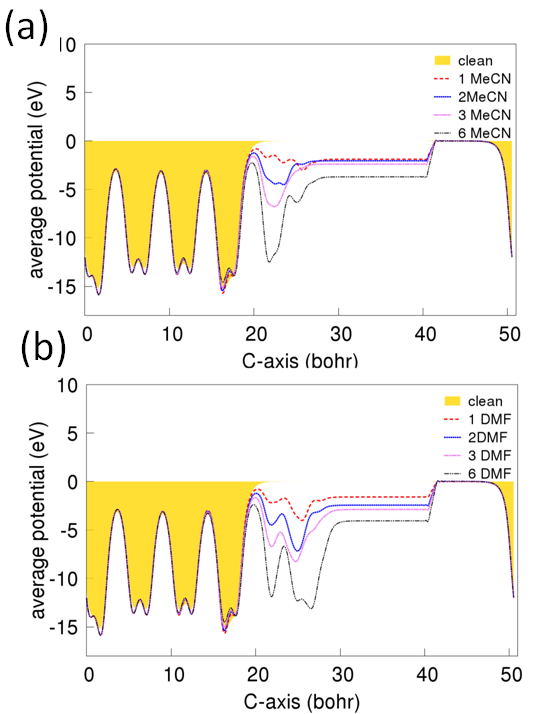}
\caption{Averaged electrostatic potential energy for (a)- MeCN/ZnO, and (b)- DMF/ZnO at different coverages. Red, blue, pink, and grey lines correspond to coverages 1, 2, 3, and 6 molecules, respectively. The yellow region specifies the shape of the potential for clean surface.}
\label{fig4}
\end{figure}

In Fig.~\ref{fig4}, we have plotted the averaged electrostatic potentials of MeCN/ZnO and DMF/ZnO for different coverages. We avoid to bring the plots for MeCN/TiO$_2$ and DMF/TiO$_2$, because of the similarity with the latter case. In the plots, we have identified the asymptotic potentials of clean surfaces as $V(-\infty)=0$. As is seen from Fig.~\ref{fig4}, the potentials towards $-\infty$ are identical and moving to the right, they start to split in the vicinity of the exposed surface. This splitting is more significant for higher coverages. Moving to the right of the figure, far from the adsorbed molecule, the potentials flatten to their asymptotic values of $V(+\infty)$. The rightmost part of the figure is the periodic image of the left. According to the above arguments, one can use the difference $[V(+\infty)-V(-\infty)]$ in the calculations of work-function for the exposed surface of semiconductor, which can be used for the level shifts in electrochemical applications.
 
\begin{figure}
\centering
\includegraphics[width=0.7\linewidth]{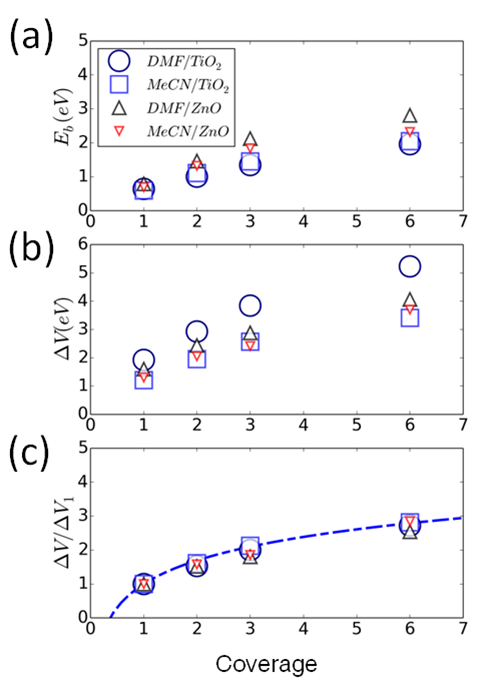}
\caption{(a)- Total binding energies, (b)- potential differences, and (c)- normalized potential differences as functions of the coverage for $X/Y$ systems with $X$=MeCN, DMF; and $Y$=TiO$_2$, ZnO.}
\label{fig5}
\end{figure}

In Fig.~\ref{fig5}, we have plotted the binding energies, potential differences, and normalized potential differences as functions of the coverage for $X/Y$ systems where $X$=MeCN, DMF; and $Y$=TiO$_2$, ZnO. As is seen from the figure, DMF adsorption leads to a greater potential difference than for MeCN, which is due to its larger dipole moment. Also we see a non-linear behavior for both binding energies and potential differences as functions of coverage. Interestingly, the values of normalized potential differences are very close to each other for different coverages, and this behavior is nicely fitted to a logarithmic function as:

\begin{equation}\label{eq7}
\frac{\Delta V_n}{\Delta V_1}=\ln(n)+1
\end{equation}  

\noindent where $\Delta V_n$ is the potential difference for coverage $n$ [See Fig.~\ref{fig5}(c)].

\begin{figure}
\centering
\includegraphics[width=0.9\linewidth]{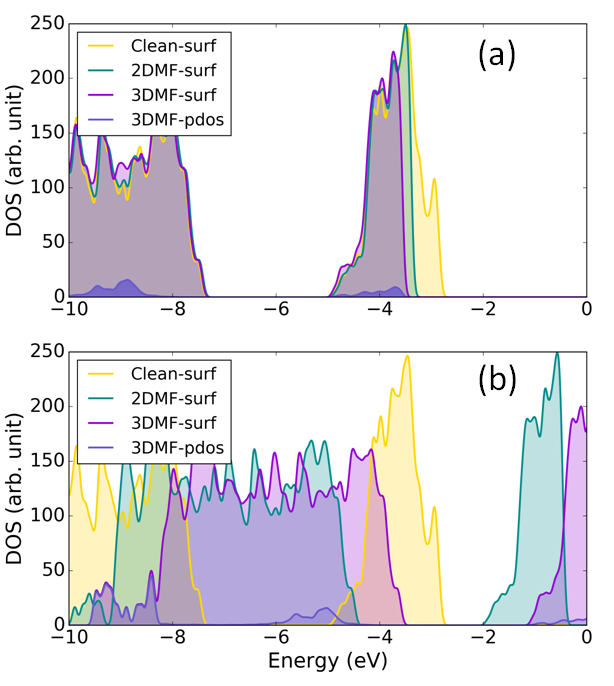}
\caption{DOS for clean (yellow) and exposed surfaces of DMF/TiO$_2$ at coverages 2 (green) and 3 (violet). Projected DOS on DMF atomic orbitals is specified as blue regions. (a)- The energy reference is taken at $V(-\infty)$, and (b)- the energy reference is taken at $V(+\infty)$.}
\label{fig6}
\end{figure}

To relate the above discussions to level alignments, we consider the density of states (DOS) for clean and exposed surfaces. Since the DMF/TiO$_2$ system has the largest value for $\Delta V$ than other $X/Y$ systems, we have calculated the DOS values of this system at two coverages and compared with that of clean surface in Fig.~\ref{fig6}. To find the absolute values of VB and CB edges, we have to refer all energies to the vacuum level. Choosing this reference energy is a crucial step to compare different systems and to investigate level shifts. The method is a common approach in level-alignment problems\cite{Kharche2014,mosconi2012solvent}.   

In the case of symmetric slabs, the asymptotic potentials are the same in both sides and can be used as a unique energy reference. However, for asymmetric slabs, as we discussed above, the asymptotic potentials at the two sides are different and we have used both of them, one at a time, as references to generate our DOS plots as in Fig.\ref{fig6}.

In Fig.~\ref{fig6}(a), the asymptotic potentials of the clean side of the slabs are taken as reference point of energy, i.e., $V(-\infty)=0$. The result is that the VB and CB edges do not change with coverage, and consequently the band gap remains unchanged. This result can be understood by the fact that, irrespective of the coverage of right side of the slab, the energy needed to extract an electron from the clean surface should be the same (the slabs are thick enough so that the surfaces at the two sides do not see each other). In addition, since the projection of DOS over atomic orbitals of DMF molecules, i.e., the dark blue small peaks in Fig.~\ref{fig6}(a), are localized far from the gap region, the adsorption does not change the DOS shapes near to the gap.

Now, if we take the asymptotic potentials of the right hand side as the reference points, we obtain the result shown in Fig.~\ref{fig6}(b). In this figure, the energy level up-shifts due to solvent molecules are clearly shown. As is seen, for higher coverage values, less energy is needed to extract an electron from the exposed surface. This case of referencing is the one that should be considered in an electrochemical system, and the values of $\Delta V$ listed in Table~\ref{tab1} can be used as a good estimate of level shifts in the presence of solvent. Once again, it should be emphasized that the energy values that we are working with, are the ones obtained from the solution of the KS equations, which are not necessarily the actual values for a system; however, using the method proposed in this work gives a good estimate for the level shifts in an electrochemical system.  

\begin{figure}
\centering
\includegraphics[width=0.9\linewidth]{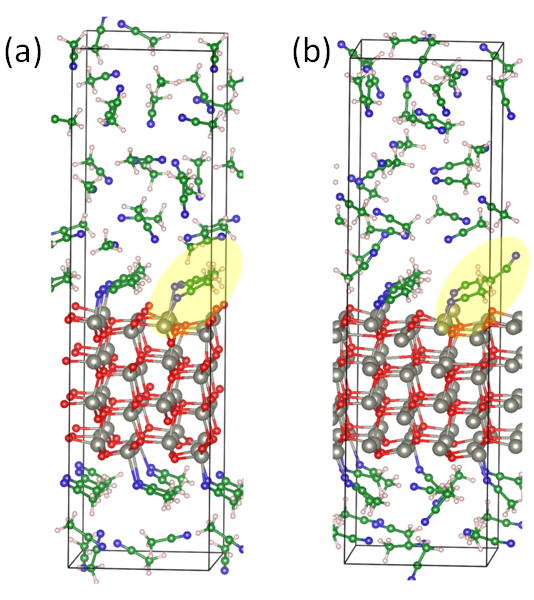}
\caption{(a)- Starting configuration, and (b)- the equilibrated configuration for a ZnO slab embedded in MeCN molecules. Yellow region highlights the detached molecule with 2.5~ps.}
\label{fig7}
\end{figure}

The results presented up to now, were based on the assumption of mono-layer adsorption in static conditions. However, in reality, the adsorbed layer interacts with other solvent molecules and there exist some thermal fluctuations at room temperatures. To have an estimation of how these effects may change our static results, we have also performed an {\it ab initio} molecular dynamics (AIMD) calculations for a ZnO slab embedded in MeCN molecules. For the starting configuration, Fig.~\ref{fig7}(a), we took a supercell containing a (ZnO)$_{48}$ slab and a number of 35 MeCN molecules inside the simulation box (corresponding to the solvent density of 0.786 g.cm$^{-3}$) out of which 12 molecules are attached (full coverage) to the proper adsorption sites of the slab surfaces (6 molecules at each side). The initial configuration for the AIMD simulations was constructed from relaxed structure of one-layer covered slab combined with classical molecular dynamics simulation of MeCN box which had been equilibrated for 100~ps at 300$^\circ$~K. Starting from this initial configuration, we have performed canonical ($NVT$) simulations for 2.5~ps using the Nose-Hoover thermostat at 300$^\circ$~K, with a time step of 0.5~fs, as implemented in SIESTA code package. 

Tracing the evolution of the system, we observe that within about 200~fs, one of the initially adsorbed molecules detaches from the surface and changes its orientation to the opposite direction, as shown in yellow highlighted region in Fig.~\ref{fig7}(b). For more details, the animated dynamics of system is included as supporting information.
To summarize, the results obtained from the molecular dynamics simulation of the system show that, the thermal effects and interaction of first-layer solvent molecules with other layers prevent the adsorption with a full coverage, and this in turn, means that the level shifts realized in actual conditions are somewhat smaller than those corresponding to full coverage, as listed in Table~\ref{tab1}. Although in this work we performed the MD simulation in a small (both in size and time) scale, the results are very informative. Extending this study for larger scale simulations is currently in progress.

\section{Conclusions}
In this study, we have used first-principles DFT calculation and {\it ab initio} MD to explore the energy-level shifts of the semiconductors TiO$_2$ and ZnO at the interfaces with MeCN and DMF solvent molecules. The DFT calculations are performed for different coverages of solvent molecules, and we have shown that the binding energies per molecule decreases with coverage. The dipole correction method is used to determine the potential difference between the clean and exposed surfaces of an asymmetric slabs. This quantity gives an estimate for the level shifts in electrochemical systems. The calculations show that some charge is injected from the adsorbed molecules into the interface, giving rise to a surface dipole layer. In the studied cases, the dipoles originating from the charge injection adds to the permanent dipole of the adsorbent molecules, leading to a higher potential difference, which in turn enhances the level shifts. Our electronic structure calculations show that the studied adsorbed molecules have negligible effects on the DOS shapes near the gaps; and looking from the exposed side of the slab, this gap shifts to the right with coverage. Finally, the dynamics of interface is studied using {\it ab initio} MD calculations for MeCN/ZnO system, and the results for 300$^\circ$~K show that in the equilibrate state some fraction of solvent molecules detach from the surface, which implies that in our static calculations we should take smaller coverages in order to mimic a realistic system.      

\appendix
\section{Charge Displacement and Dipole Moments} \label{appsec1}
The total charge densities of an isolated molecule and an isolated surface (slab), in their equilibrium ground states, are given respectively, by: 

\begin{equation}\label{appeq1}
\rho_{\rm mol}({\bf r})= +e\sum_{\alpha} Z_\alpha \delta ({\bf r}-{\bf R_\alpha})-e n_{\rm mol}({\bf r};\{{\bf R_\alpha}\})
\end{equation}

\begin{equation}\label{appeq2}
\rho_{\rm sur}({\bf r})= +e\sum_{\beta} Z_\beta \delta ({\bf r}-{\bf R_\beta})-e n_{\rm sur}({\bf r};\{{\bf R_\beta}\})
\end{equation}
where, $e$ is the magnitude of an electronic charge, and the sets $\{{\bf R_\alpha}\}$ and $\{{\bf R_\beta}\}$ determine the ionic position vectors of the molecule and slab, respectively; while $n_{\rm mol}({\bf r};\{{\bf R_\alpha}\})$ and $n_{\rm sur}({\bf r};\{{\bf R_\beta}\})$ are the ground state electronic number densities of molecule and slab, respectively, which are obtained by the solutions of the KS equations of DFT for the given external parameters of the ionic positions.

When the molecule is adsorbed on the surface, the set of equilibrium ionic positions in the combined system change to $\{{\bf R^\prime_\alpha},{\bf R^\prime_\beta}\}$, and the total charge density of the combined system, in its ground state, is given by:

\begin{widetext}
\begin{equation}\label{appeq3}
\rho_{\rm comb}({\bf r})= +e\sum_{\alpha} Z_\alpha \delta ({\bf r}-{\bf R^\prime_\alpha}) +e\sum_{\beta} Z_\beta \delta ({\bf r}-{\bf R^\prime_\beta})          -e n_{\rm comb}({\bf r};\{{\bf R^\prime_\alpha},{\bf R^\prime_\beta} \})
\end{equation}
\end{widetext}
Now, to obtain the charge difference, Eq.~(\ref{eq3}), the total charge densities of the isolated molecule and surface should be calculated for the ground states with external parameters $\{{\bf R^\prime_\alpha}\}$ and $\{{\bf R^\prime_\beta} \}$, respectively, so that the contributions from the ionic charges cancel out:

\begin{widetext}
\begin{equation}\label{appeq4}
\Delta\rho({\bf r}) = -e n_{\rm comb}({\bf r};\{{\bf R^\prime_\alpha},{\bf R^\prime_\beta} \}) + e n_{\rm mol}({\bf r};\{{\bf R^\prime_\alpha}\}) + e n_{\rm sur}({\bf r};\{{\bf R^\prime_\beta}\})
\end{equation}    
\end{widetext}

\noindent where the electronic charge density is defined as the multiplication of the electronic charge and the number density.

To obtain the dipole moment of the combined system, $\mu_{\rm comb}$, we multilpy by {\bf r} both sides of Eq.~(\ref{appeq4}) and integrate over the whole space:

\begin{widetext}
\begin{eqnarray}
\mu_{\rm chem}\equiv \int{\bf r}\;\Delta\rho({\bf r})\; d{\bf r} &=& -e \int {\bf r}\; n_{\rm comb}({\bf r};\{{\bf R^\prime_\alpha},{\bf R^\prime_\beta} \}) \; d{\bf r} \\
 && \nonumber + e \int {\bf r}\; n_{\rm mol}({\bf r};\{{\bf R^\prime_\alpha}\}) \; d{\bf r} + e \int {\bf r}\; n_{\rm sur}({\bf r};\{{\bf R^\prime_\beta}\}) \; d{\bf r}
\end{eqnarray}
\end{widetext}

\noindent Rearranging the terms and by the definition of the electric dipole moment as the integral over space of the product of {\bf r} and charge density, we obtain Eq.~(\ref{eq4}).  

\begin{acknowledgments} 
This work is part of research program in School of Physics and Accelerators, NSTRI, AEOI.  
\end{acknowledgments}

\bibliography{payami-arxiv-revised}

\end{document}